\documentclass[sigconf]{acmart}

\usepackage{booktabs} 
\usepackage{graphicx}
\usepackage{xcolor}
\usepackage{amssymb}
\usepackage[linesnumbered,ruled]{algorithm2e}
\usepackage{algorithmicx}
\usepackage{amsmath}
\usepackage{subfig}
\setcopyright{rightsretained}

\acmDOI{10.475/123_4}

\acmISBN{123-4567-24-567/08/06}

\begin{document}
\title{STWalk: Learning Trajectory Representations in Temporal Graphs}

 \author{Supriya Pandhre}
 \affiliation{%
   \institution{Indian Institute of Technology Hyderabad}
   \city{Hyderabad, India}
 }
 \email{cs15mtech11016@iith.ac.in}

 \author{Himangi Mittal}
 \affiliation{%
   \institution{Jaypee Institute of Information Technology}
   \city{Noida, India}
 }
 \email{himangimittal@gmail.com}
 
 \author{Manish Gupta}
 \affiliation{%
   \institution{Microsoft}
   \city{Hyderabad, India}
 }
 \email{gmanish@microsoft.com}

 \author{Vineeth N Balasubramanian}
 \affiliation{%
   \institution{Indian Institute of Technology Hyderabad}
   \city{Hyderabad, India}
 }
 \email{vineethnb@iith.ac.in}

\newcommand{\STE}{STWalk}
\newcommand{\STEone}{STWalk1}
\newcommand{\STEtwo}{STWalk2}

\begin{abstract}
Analyzing the temporal behavior of nodes in time-varying graphs is useful for many applications such as  targeted advertising, community evolution and outlier detection. In this paper, we present a novel approach, \STE, for learning trajectory representations of nodes in temporal graphs. The proposed framework makes use of structural properties of graphs at current and previous time-steps to learn effective node trajectory representations. \STE\ performs random walks on a graph at a given time step (called \textit{space-walk}) as well as on graphs from past time-steps (called \textit{time-walk}) to capture the spatio-temporal behavior of nodes. We propose two variants of \STE\ to learn trajectory representations. In one algorithm, we perform space-walk and time-walk as part of a single step. In the other variant, we perform space-walk and time-walk separately and combine the learned representations to get the final trajectory embedding. Extensive experiments on three real-world temporal graph datasets validate the effectiveness of the learned representations when compared to three baseline methods. We also show the goodness of the learned trajectory embeddings for change point detection, as well as demonstrate that arithmetic operations on these trajectory representations yield interesting and interpretable results.
\end{abstract}

\copyrightyear{2018}
\acmYear{2018}
\setcopyright{acmcopyright}
\acmConference[CoDS-COMAD '18]{The ACM India Joint International Conference on Data Science \& Management of Data}{January 11--13, 2018}{Goa, India}
\acmBooktitle{CoDS-COMAD '18: The ACM India Joint International Conference on Data Science \& Management of Data, January 11--13, 2018, Goa, India}
\acmPrice{15.00}
\acmDOI{10.1145/3152494.3152512}
\acmISBN{978-1-4503-6341-9/18/01}


%
%



\maketitle

\section{Introduction}
The rising use of social networks and various other sensor networks in real-world applications ranging from politics to healthcare has made computational analysis of graphs a very important area of research today. The use of machine learning for various graph analysis tasks such as community detection ~\cite{malliaros2013clustering}, link analysis ~\cite{getoor2005link}, node classification ~\cite{aggarwal2011introduction} and anomaly detection ~\cite{akoglu2015graph} has exponentially increased over the last few years, considering the direct impact of the success of these methods on business outcomes. A wide variety of methods have been proposed in the aforementioned areas over the last few years. However, a large part of the efforts so far has focused on static graphs. Platforms such as Facebook, Twitter and LinkedIn correspond to graphs that change on a daily basis; however, the algorithms that analyze such graphs are primarily static, i.e., they consider a graph at a given snapshot of time for analysis. In this work, we seek to focus on temporal graphs, and on learning representations of node trajectories in such graphs.

Study of user-user interactions over time plays an important role in many applications, including user classification for targeted advertising or link prediction to suggest new connections on social networking sites. One of the crucial part of such applications is analyzing change in users' behavior over time. Aggarwal and Subbian ~\cite{aggarwal2014evolutionary} present a survey of methods that have been proposed so far to study temporal graphs, and categorize such methods into two kinds: \textit{maintenance methods} that focus on adapting older models to a newer version of the same graph, or \textit{analytical evolution analysis}, where the objective is to analyze the network as it changes itself. We propose to address the latter in this work. Recent efforts for analysis of temporal graphs have largely extended earlier methods proposed on static graphs, such as spectral methods or those that study statistical properties. In this work, we build on the recent emphasis on learning appropriate representations of data to temporal graphs. In particular, we propose SpaceTimeWalk (\textit{\STE}), which learns representations of node trajectories in temporal graphs. To the best of our knowledge, this is the first such effort.

\textit{\STE} is an unsupervised trajectory learning algorithm that embeds the spatial and temporal characteristics of nodes over a given time window. The algorithm carries out a random walk in a current graph, resulting in a \textit{space-walk}, as well as in graphs at different time steps, called \textit{time-walk}. The traversed paths are then considered as sentences, with nodes as words from a vocabulary. The SkipGram ~\cite{mikolov2013efficient} network
is then employed to learn the latent representations of node trajectories such that it maximizes the probability of co-occurrence of two nodes within the specified window size. We perform extensive experimentation on three real-world time-varying networks. Performance on these datasets show the effectiveness of our framework in dealing with different kinds of temporal networks. We also study the usefulness of these representations through visualization, as well as complementary tasks such as change point detection.

The remainder of this paper is organized as follows. Section \ref{sec_related_work} reviews related work in relevant areas, highlighting the need for the proposed work. Section \ref{sec:algomethod} presents the mathematical formulation of the problem, and the proposed methods are presented in Sections \ref{subsec:algo1} and \ref{subsec:algo2}. The experimental results that study the performance of the proposed methods are described in Section \ref{sec:exp}, and Section \ref{sec:conc} concludes the work with pointers to future directions.

\section{Related Work}
\label{sec_related_work}
Our work is most closely related to two sub-areas of graph analysis that have garnered increased interest in the last few years: network representation learning and temporal graph analysis. We review methods in both of these sub-areas below.

\subsection{Network Representation Learning}

In recent years, there have been a few approaches proposed for learning representations of networks, in particular, learning graph node embeddings. Many of the recent efforts have been inspired by the recent wave of deep learning methods that have demonstrated impressive success in learning representations of other kinds of data such as images, speech and text. DeepWalk~\cite{Perozzi:2014:DOL:2623330.2623732}, one of the earliest efforts in this regard, combined a random walk-based method with a SkipGram network (traditionally used in Natural Language Processing) to learn node representations. Node2vec~\cite{grover2016node2vec} extended DeepWalk by employing biased random walks to learn node embeddings. In particular, Node2Vec precomputes a matrix, which is similar to a transition probability matrix, that assigns a probability of transition in a random walk, and is based on two hyper-parameters that control whether the walk will stay within a close neighborhood of the node or it will move farther away from that node.

Tang et al. (LINE)~\cite{tang2015line} computed node embeddings using a function of the probability of observing first-order and second-order nodes together in a random walk within a certain window size. This method then uses Asynchronous Stochastic Gradient Descent to solve the optimization problem. GraRep~\cite{cao2015grarep} employed a Singular Value Decomposition-based method for dimensionality reduction and learning graph representations. Cao et al.~\cite{cao2016deep} proposed a stacked denoising autoencoder for learning the vertex representations in graphs, where they used random surfer model to learn the structural properties of graph calculate the positive pointwise mutual information and then employ denoising autoencoder to reduce the dimensions of learned node features. There have also been efforts, such as~\cite{chang2015heterogeneous}, which have used deep neural network-based models for learning embeddings in heterogeneous networks. The main goal of the paper~\cite{chang2015heterogeneous} was to consider information from multiple sources in dynamic network and learn unified representation. A more detailed review on network embedding techniques can be found in~\cite{goyal2017graph}. However, all the aforementioned methods work only with static graphs. In this work, we focus on learning representations of node trajectories in dynamic graphs. Dynamic graphs are evolving by nature, and it is important to capture the addition and deletion of nodes and edges, while capturing the behavior of a node over time in a learnt representation.

\subsection{Temporal Graph Analysis}
Related to the other focal subarea for this work, time-varying graphs have been studied for applications such as node classification~\cite{aggarwal2011node,hamilton2017inductive,li2017attributed}, link prediction~\cite{sarkar2012nonparametric}, community evolution~\cite{tang2008community} and outlier detection~\cite{gupta2014outlier} in recent years. In one of the earlier works in this area, Tang et al.~\cite{tang2008community} proposed a spectral clustering framework for studying the evolution of communities in time varying multi-mode networks. Aggarwal et al.~\cite{aggarwal2011node} proposed a node classification model that used network structure and node attributes from time-varying graphs. The model was specifically designed for textual node attributes.  In~\cite{sarkar2012nonparametric}, a non-parametric model was proposed for link prediction in a series of network snapshots. More recently, Li et al.~\cite{li2017attributed} presented a framework that used node attributes for learning node embeddings. The model first learns separate embeddings from network structure and node attributes respectively. The node embeddings are then combined using matrix perturbation theory. Another recent work, GraphSAGE~\cite{hamilton2017inductive}, proposed an inductive algorithm that generates node embeddings using sampling and aggregating local neighborhood information and learning a function to generate node embeddings of unseen data. For more such methods, Aggarwal et al.~\cite{aggarwal2014evolutionary} provide a comprehensive survey on methods for temporal graph analysis.

While there have been a few efforts on analysis of time-varying graphs, all the aforementioned algorithms enforce the data to have node attributes. Many real-world graph data, however, may not provide node features either because they are not available publicly for use or they are missing (for instance, it is optional to fill age or location while creating a user profile on Twitter). In this work, we seek to propose a new framework, SpaceTimeWalk (\textit{\STE}), that seeks to utilize only the structural properties of time-varying graphs to learn compact low-dimensional embeddings capturing spatio-temporal properties of node trajectories.

We now present the proposed \textit{\STE} algorithm.
\section{Proposed method: \STE}
\label{sec:algomethod}
Learning the change in behavior of nodes in a temporal graph requires us to consider the information from the graph in the current time step, as well as graphs from previous time steps. Using the na\"{\i}ve way of examining graphs at each time-step separately will fail to capture the correlation information that exists between two graphs at consecutive time steps. Hence, we propose \STE\, a random walk based framework that learns rich trajectory representations by capturing changes in node behavior across a given time window. We hence define the following to formulate our solution.

\begin{itemize}
\item Given a graph $G=(V,E)$ at $T$ different time steps $\{G_1,G_2,G_3,\cdots,G_T\}$ having $N$ number of nodes at each time step and varying number of edges $E_t, t \in \{1,\cdots,T\}$, where $E_t$ represents the edges in $G_t$.
\item Let $W$ denote the adjacency tensor of size $N \times N \times T$, representing the adjacency matrix of the graph at different time stamps.
\end{itemize}

Our goal is to learn representations $\Phi$ that maps a given node to a $d$-dimensional representation. Let $N_t(u) \subset V$ denote the set of neighbors of node $u$ in graph $G_t$. For a node $u$ in graph $G_t$, the representation $\Phi_t(u)$ will be learned in such a way that, it maximizes the probability of observing the neighbors of $u$ in graph $G_t$, viz. $N_t(u)$, and also the representation of $u$ at previous time stamps $\Phi_{t-\tau}(u)$ where $\tau \in \{1,\cdots, t-1\}$. Equation ~\eqref{eqn:01} below formalizes this construction. Our objective hence is to obtain a $d$-dimensional representation for a node $u$ at time $t$ that maximizes the following log probability:

\begin{equation}
\label{eqn:01}
\underset{\phi_t}{\max} \underset{{u\in V}}{\sum} \log \text{Pr}\left(N_t(u),\ \phi_{t-\tau}(u)\ |\ \phi_t(u))\right)
\end{equation}
 
\[\text{where } \tau \in \{1,\cdots,t-1\}\]

We assume that the neighborhood nodes in $N_t(u)$ and past representations $\phi_{t-\tau}(u)$ are independent of each other, conditioned on the current representation of $u$, $\phi_t(u)$. Hence, we write:

\begin{equation}
\label{eqn:02}
\begin{split}
&\log \left(\text{Pr}(N_t(u),\phi_{t-\tau}(u)|\phi_t(u))\right)\\
=&\log \left( \underset{{n_i\in N_t(u)}}{\prod} \text{Pr}(n_i|\phi_t(u)) \right) + \log \left(\text{Pr}(\phi_{t-\tau}(u)|\phi_t(u))\right)
\end{split}
\end{equation}

Learning representations using random walk has proved to measure better graph proximity, and thereby improving the performance ~\cite{goyal2017graph} ~\cite{hamilton2017representation}. Hence, we use random walk to learn the conditional probability of observing a node $n_i$ given the learned representation $\phi_t(u)$ defined  as follows:
\begin{equation}
\label{equ:03}
\text{Pr}(n_i|\phi_t(u)) = \frac{\exp(\phi_t(n_i)\phi_t(u))}{\underset{v \in N_t(u)}{\sum}\exp(\phi_t(v)\phi_t(u))}
\end{equation}

Therefore, to learn the trajectory representation, we maximize:
\begin{equation}
\label{eqn:04}
\begin{split}
\underset{\phi_t}{\max} \underset{{u\in V}}{\sum} \log \left( \underset{{n_i\in N_t(u)}}{\prod}  \frac{\exp(\phi_t(n_i)\phi_t(u))}{\underset{v \in N_t(u)}{\sum}\exp(\phi_t(v)\phi_t(u))} \right)\\
+ \log \left(\text{Pr}(\phi_{t-\tau}(u)|\phi_t(u))\right)
\end{split}
\end{equation}

\noindent While the above problem appears to be learning only the representation of a node at time $t$, it actually represents a representation of a node $t$ considering its history over the graphs at the previous time steps, and hence, the term \textit{node trajectory representation} in this work.

To solve the maximization problem defined in Equation \ref{eqn:04}, we propose two approaches. The first approach solves equation \ref{eqn:04} by considering it as one single expression and learning a trajectory representation by maximizing the co-occurrence of neighborhood nodes at time $t$ and in previous $t-\tau$ time-steps together. In the second approach, we learn the trajectory representation by solving two sub-problems separately: (i) maximizing the co-occurrence probability of a node and its neighbors at time $t$, and (ii) maximizing the co-occurrence probability of a node and its neighbors at preceding $t-\tau$ time-steps. The representations learned using these steps are then combined to form the final trajectory representation. Sections \ref{subsec:algo1} and \ref{subsec:algo2} describe each of the proposed approaches.

\subsection{\STEone}\label{subsec:algo1}
We describe the first approach, which we call \STEone, in this section. For learning the trajectory representations of nodes in temporal graphs, here, we take into account the neighbors of a node at time steps $t$, $t-1$,..$t-\tau$ together and maximize the probability of observing these nodes together.

\begin{figure}
\includegraphics[width=8cm,height=5cm]{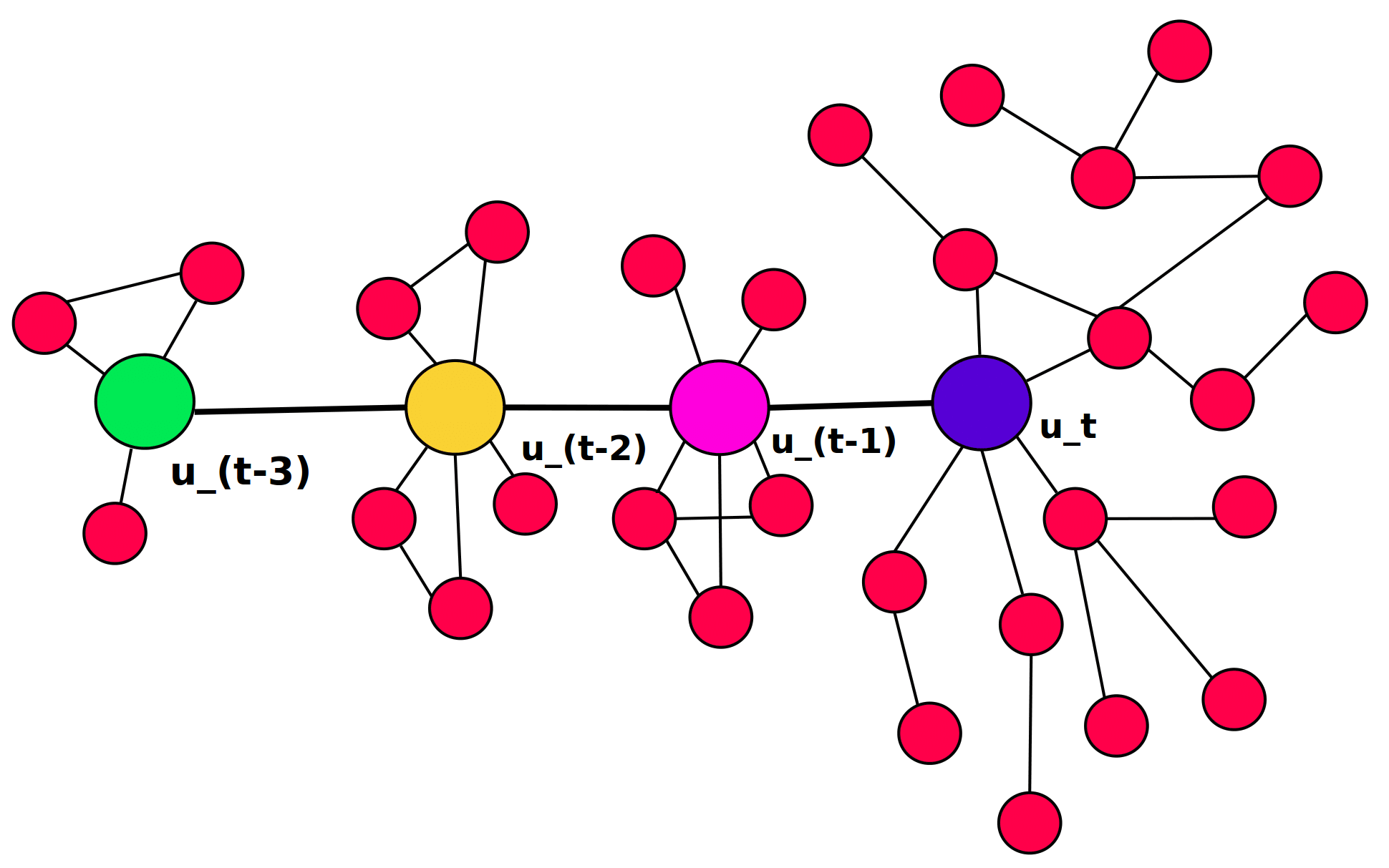}
\caption{Example graph of node u\_t generated by Algorithm createSpaceTimeEmb \ref{alg:createSpaceTimeGraph}. The node $u_t$ represents node in current graph and nodes $u\_(t-1)$, $u\_(t-2)$ and $u\_(t-3)$ indicate its past self-nodes. To learn trajectory representation for $u\_t$, \STEone\ algorithm considers the influence of nodes in current time step as well as effect from the past.}
\label{fig1:impl-1}
\end{figure}

In \STEone, to consider the neighbors of a node $u$ from the current graph and graphs at previous time steps, we construct a Space-Time Graph. The space-time graph contains all nodes from graph at time $t$ and it has a special temporal edge that links the node to itself in the graphs at previous time steps. Figure \ref{fig1:impl-1} illustrates this with an example of a space-time graph considered while learning the trajectory representation for node at time $t$, $u_t$. In particular, the figure shows the temporal edge (shown as a bold line) between $u_t$ and past self-nodes $u_{t-1}$, $u_{t-2}$ and $u_{t-3}$. The trajectory representation thus learned for $u_t$ will hence be influenced by all the nodes in the Space-Time graph, as shown in Figure \ref{fig1:impl-1}, viz. nodes from the present graph as well as its preceding self nodes and their first-level neighbors. Given a node and the window size, over which we want to learn the trajectory, Algorithm \ref{alg:createSpaceTimeGraph} describes the steps to create a Space-Time Graph starting from the given node.

Now, to learn the representations of nodes in the Space-Time graph, we perform random walks on it. If two nodes share many edges or neighbors, they will be visited more often in many random walks, indicating that these two nodes share a similar graph structure. Hence, the representation learned for these two nodes should be close to each other in the embedding space. This is analogous to the concept from Natural Language Processing (NLP) that if two words co-occur in many sentences, this indicates that they represent a similar context and hence their word vectors are nearby each other in word embedding space. To learn such a embedding that captures the relationship of a word with other co-occurring words in a window, Mikolov et al.~\cite{mikolov2013efficient} presented the SkipGram network. We use a similar SkipGram algorithm (given in Algorithm \ref{alg:skipgram}) to learn the node embeddings from random walks of a graph, similar to DeepWalk~\cite{Perozzi:2014:DOL:2623330.2623732}. For more details of the SkipGram network, we request the interested reader to refer to~\cite{mikolov2013efficient}.

For each node in the Space-Time graph, we perform $\rho$ number of random walks each of length $\mathcal{L}$.  
The problem of learning the trajectory representation in such a way that it maximizes probability of co-occurrence of present and past nodes is reduced to maximizing the co-occurrence of nodes in a random walk within a vocabulary window $\mathcal{W}_v$. Hence, we use the SkipGram algorithm \ref{alg:skipgram} to learn the trajectory representations for nodes. (In particular, we use the SkipGram network, as proposed in \cite{mikolov2013efficient} to implement the SkipGram algorithm.) For each node in a graph $G_t$, two representations will be learned. One representation corresponds to the node embedding when it occurs in the context of other nodes, and the other representation which is learned for the node itself. The latter one is used as the node trajectory representation. The overall algorithm \STEone\  is illustrated in Algorithm~\ref{alg:algo1}.

\begin{algorithm}[h]
\SetAlgoNoLine
\KwIn{ Set of graphs $\mathcal{G}$:\{$G_t,G_{t-1},\cdots,G_{t-\tau}$\}, Time window size $\tau$, starting node $startNode$ at time $t$}
\KwOut{$G_{spacetime}$}
$G_{spacetime}\ = G_t$\\
\For{each time step $i \in \{0,\cdots,\tau\}$}{
   $pastNode$ = $startNode$ at $G_{t-i}$\\
   Create $pastNodeSubGraph$= the subgraph of $pastNode$ and its direct neighbors from $G_{t-i}$\\
   Merge $G_{spacetime}$ and $pastNodeSubGraph$ to obtain updated $G_{spacetime}$\\
   $startNode$ = $pastNode$
  }
\caption{createSpaceTimeGraph}
\label{alg:createSpaceTimeGraph}
\end{algorithm}


\begin{algorithm}
\SetAlgoNoLine
\KwIn{Input representation: $\Phi$, List of nodes in a random walk: $walk$, windowSize: $\mathcal{W}_v$}
\KwOut{Updated representation: $\Phi$}
\For{$n \in walk$}{
	\For{$n_i\in$ walk[$n-\mathcal{W}_v$, $n+\mathcal{W}_v$]}{    	
      	$J(\Phi[n]) = -\log$ $\text{Pr}(\Phi[n_i]|\Phi[n])$ \\    
        $\Phi[n] = \Phi[n] - \frac{\partial J(\Phi[n])}{\partial \Phi[n]}$
      }
  }
\caption{SkipGram}
\label{alg:skipgram}
\end{algorithm}


\begin{algorithm}[h]
\SetAlgoNoLine
\KwIn{Given node $n \in [1,N]$,\\ 
time step $t \in [1,T]$,\\
length of random walk: $\mathcal{L}$,\\
temporal window size: $\tau$,\\
vocabulary window size: $\mathcal{W}_v$,\\
Set of graphs $\mathcal{G}$:\{$G_t,G_{t-1},\cdots,G_{t-\tau}$\},
size of embedding: $d$,\\
number of restarts (starts at same node): $\rho$}
\KwOut{Updated  $n^{th}$ row of matrix of $\Phi_t$ of size: $N \times d$}
Initialize $\Phi_t[n] \in \mathbb{R}^{d}$ \\
\For{i=0 to $\rho$}{
      	new\_graph = createSpaceTimeGraph($\mathcal{G}$,$\tau$,$n$)\\
        spaceTimeWalk = randomWalk$(\text{new\_graph}, $n$,\mathcal{L})$;\\
        
        $\Phi_t[n]$ = SkipGram ($\Phi_t$, spaceTimeWalk, $\mathcal{W}_v$);
  }
\caption{\STEone}
\label{alg:algo1}
\end{algorithm}

\subsection{\STEtwo}\label{subsec:algo2}

In this section, we detail the second approach for learning the trajectory representation of nodes in a temporal graph. We solve the problem of learning representations of node trajectories, such that these representations are influenced by present and past neighbors, by dividing it into two sub-problems. We learn the spatial representation from the current graph, $G_t$, and the temporal representation from past graphs $G_{t-1},\cdots,G_{t-\tau}$, separately. The final trajectory representation is obtained by combining the spatial and temporal representations of a given node.

The \STEtwo\ algorithm first performs a random walk in the current time step graph $G_t$, called as \textit{SpaceWalk}, and learns the spatial representation for each node using the SkipGram algorithm. $\Phi^{space}_{u,:}$ corresponds to the learned spatial representation of node $u$. To learn the temporal representation of a node, \STEtwo\ constructs $G_{new}$, which contains only one node from current graph and neighbors from past graphs. For example, in order to construct such a $G_{new}$ in Figure \ref{fig1:impl-1}, we remove all neighbors from current time-step graph of node $u_t$. The remaining graph will have $u_t$ linked only to past subgraphs of $u_{t-1}$, $u_{t-2}$ and $u_{t-3}$. This will be used as $G_{new}$ to learn the temporal representation (in a manner similar to learning the spatial representation above).

The random walks on $G_{new}$ capture the temporal relationship between the current node and nodes from its past graphs. $\Phi^{time}_{u,:}$ corresponds to the learned temporal representation of node $u$. Finally, the two learned embeddings are combined to get trajectory representation of nodes. Algorithm \ref{alg:algo2} summarizes this approach. Line 15 of Algorithm \ref{alg:algo2} states that the final trajectory representation is a function of spatial and temporal representations. In this work, we found vector addition to work very well in practice in the experiments.

Thus, in \STEtwo, we ensure that the trajectory representation of the current node $u_t$ has influence from past nodes by learning the temporal representation explicitly. Learning the temporal representation separately forces the random walk to explore the past graphs of the same node and capture the temporal information between the current node and its past. Therefore, \STEtwo\ is able to learn richer trajectory embeddings than \STEone\, because in \STEone\, we consider the present and past neighbors together, giving the random walks an option of skipping few nodes from past graphs. Experimental results in Table \ref{table:accuracy} validate that the trajectory embeddings learned by \STEtwo\ perform better than embeddings learned by \STEone. We now present details of our experiments.


\begin{algorithm}
\SetAlgoNoLine
\KwIn{Given node $n \in [1,N]$,\\
time step $t \in [1,T]$
Set of graphs $\mathcal{G}$:\{$G_t,G_{t-1},\cdots,G_{t-\tau}$\},\\
length of spatial random walk : $\mathcal{L}_\alpha$,\\
length of temporal random walk : $\mathcal{L}_\beta$,\\
temporal window size: $\tau$,\\
vocabulary window size: $\mathcal{W}_v$,\\
size of embedding: $d$,\\
number of restarts (starts from same node): $\rho$}
\KwOut{Updated  $n^{th}$ row of matrix of $\Phi_t$ of size: $N \times d$}
Initialize $\Phi_t[n] \in \mathbb{R}^{d}$  \\
Initialize $\Phi^{space}$, trajectory representation matrix of nodes only at $G_t$\\
Initialize $\Phi^{time}$, trajectory representation matrix of past nodes at $G_{t-1},\cdots,G_{t-\tau}$ \\ 
\For{i=0 to $\rho$}{
  //for space walk\\
  spaceWalk = randomWalk($G_t$, $n$,$\mathcal{L}_\alpha$)\\
  $\Phi^{space}[n]$ = SkipGram($\Phi^{space}$, $spaceWalk$,$\mathcal{W}_v$)\\
  //for time walk \\
  Construct $G_{new}$ as described in Sec \ref{subsec:algo2}\\
  timeWalk = randomWalk($G_{new}$, $n$,$\mathcal{L}_\beta$)\\
  $\Phi^{time}[n]$ = SkipGram($\Phi^{time}$, $timeWalk$,$\mathcal{W}_v$)\\  
  //final trajectory embedding\\
  $\Phi_t[n]$ = $f(\Phi^{space}[n],\Phi^{time}[n])$
  }
\caption{\STEtwo}
\label{alg:algo2}
\end{algorithm}

\begin{table}[h]
\begin{tabular*}{\columnwidth}{|p{1.5cm}|p{1.2cm}|p{1.1cm}|p{1.5cm}|p{1.4cm}|}
\hline
Datasets&No. of nodes&No. of edges&Duration&No. of Trajectories\\
\hline
\hline
DBLP&118,030&288,634&1969-2016&15,400\\
\hline
EPINION&21,575&2,590,798&Mar 2002-Apr 2011&21,575\\
\hline
CIAO&2249&27,209&Sep 2001-Mar 2011&1800\\
\hline
\end{tabular*}
\caption{Statistics of datasets: number of nodes, number of edges, the duration of data considered for analysis and number of node trajectories whose class labels are known. This is further described in Sections \ref{subsec_datasets} and \ref{subsec:classi}.}
\label{table:statistics}
\end{table}

\begin{table*}[h]
\centering
\begin{tabular}{|p{0.7cm}|p{0.7cm}|p{0.7cm}|p{0.7cm}|p{0.7cm}|p{0.7cm}|p{0.7cm}|}
\hline
\multicolumn{2}{|c|}{ }&\multicolumn{5}{|c|}{Methods}\\
\hline
\multicolumn{2}{|c|}{ }&\multicolumn{1}{|c|}{Node PageRank}&\multicolumn{1}{|c|}{Avg DeepWalk}&\multicolumn{1}{|c|}{LSTM-AE}&\multicolumn{1}{|c|}{\STEone}&\multicolumn{1}{|c|}{\STEtwo}\\
\hline
Emb Dim&Win Size&Acc (\%)&Acc (\%)&Acc (\%)&Acc (\%)&Acc (\%)\\
\hline
\hline
\multicolumn{2}{|c|}{ }&\multicolumn{5}{|c|}{DBLP}\\
\hline
\hline
64&5&70.42&72.61&73.89&77.66&\textbf{78.74}\\
\hline
64&10&61.94&70.21&73.31&70.43&\textbf{77.53}\\
\hline
128&10&74.53&71.53&74.32&\textbf{83.72}&78.25\\
\hline
\multicolumn{2}{|c|}{ }&\multicolumn{5}{|c|}{EPINION}\\
\hline
\hline
64&5&50.03&51.44&50.08&51.82&\textbf{53.10}\\
\hline
64&10&53.24&52.21&50.50&55.51&\textbf{56.70}\\
\hline
128&10&57.36&53.79&52.70&59.03&\textbf{62.03}\\
\hline
\multicolumn{2}{|c|}{ }&\multicolumn{5}{|c|}{CIAO}\\
\hline
\hline
64&5&51.94&52.02&54.17&56.31&\textbf{57.27}\\
\hline
64&10&25.56&24.65&43.95&58.17&\textbf{62.20}\\
\hline
128&10&32.03&53.15&53.80&57.15&\textbf{60.29}\\
\hline
\end{tabular}
\caption{Trajectory classification results on the DBLP, EPINION and CIAO datasets. \STEone \ and \STEtwo\ are the proposed algorithms, while Node PageRank, Avg DeepWalk and LSTM-AE are the baseline methods. Column `Emb Dim' indicates the dimension of the trajectory representation, while `Win Size' indicates the size of time window considered for learning trajectory representations.}
\label{table:accuracy}
\end{table*}

\section{Experiments}\label{sec:exp}
We discuss details of our datasets in Section~\ref{subsec_datasets}. 
We evaluated the proposed methods primarily on the task of trajectory classification. We also extended this to study the goodness of the learned trajectory representations through visualization as well as a second task, change point detection. Their performance is compared with three baseline methods, which are described in Section \ref{subsec:baseline}. Section \ref{subsec:classi} and \ref{subsec:change} provide the details of the experimental setup and results for the trajectory classification and change point detection tasks respectively. The code and datasets used for experiments are available on github\footnote{https://github.com/supriya-pandhre/STWalk}.

\subsection{Datasets}
\label{subsec_datasets}
We use three real-world datasets with temporal graphs for the experiments in this work. Each of these datasets is described below.\\

\noindent \textbf{DBLP Dataset:} The DBLP\footnote{https://www.informatik.uni-trier.de/~ley/db/} dataset contains bibliographic information about a large collection of computer science publications. This is a dataset commonly used in graph analysis. We have considered a subset of publications which fall under four areas of research: Database, Artificial Intelligence, Data Mining and Information Retrieval. We construct graphs from the this subset of DBLP dataset by considering authors as nodes and connecting two nodes if the corresponding authors have published a paper together. We constructed 45 such annual graphs starting from year 1969 till 2016, excluding years 1970, 1972 and 1974 (whose data was not available on the DBLP website).\\

\noindent \textbf{Epinion Dataset:} Epinion\footnote{Website: http://www.epinions.com.} is a popular product review site where users write critical reviews about products from various categories. The Epinion dataset\footnote{http://www.cse.msu.edu/~tangjili/trust.html} contains review details such as user ids, product ids, category ids, time-stamp when the ratings were created, along with few other fields. This dataset has been used in earlier work such as ~\cite{tang-etal12a} and ~\cite{tang-etal12b}. We create a dynamic network from this dataset by considering users as nodes and inducing an edge between two users if they have rated products from the same category. We created 110 such graphs from monthly data starting from March 2002 until April 2011.\\

\noindent \textbf{Ciao Dataset:} Ciao\footnote{http://www.ciao.co.uk} is another popular product review site. This dataset\footnote{http://www.cse.msu.edu/~tangjili/trust.html} contains fields similar to those in Epinion: user ids, product ids, category ids, and time-stamps when the ratings were created. The temporal graph for the Ciao dataset also contains user-ids as nodes and edges between nodes when the corresponding users review products from the same category. We created 115 such graphs from monthly data starting from September 2001 until March 2011.\\

\noindent More details of each of these datasets are shown in Table \ref{table:statistics}.

\subsection{Baseline Methods}
\label{subsec:baseline}
We compare the performance of our proposed methods against the following baseline algorithms.

\noindent \textbf{Node PageRank:} This method defines a  simple approach to use graph structure to study changes in behavior of nodes over time, viz. node trajectories. In this method, we compute the PageRank value of nodes at each time step of a graph within a window. The vector containing these PageRank values at different time steps is used as a trajectory representation for the tasks evaluated in this work.\\

\noindent \textbf{Avg DeepWalk}: DeepWalk \cite{Perozzi:2014:DOL:2623330.2623732} uses a truncated random walk to represent each node in a static graph. To compare performance of \STE \ against DeepWalk, we run the DeepWalk algorithm at each time step. The average of these node embeddings is considered as trajectory representation of the node across time.\\

\noindent \textbf{LSTM-AE}: Inspired by the work in ~\cite{srivastava2015unsupervised} to learn representations of video sequences using an unsupervised approach, we use a similar approach for learning node trajectories in a temporal graph. DeepWalk ~\cite{Perozzi:2014:DOL:2623330.2623732} is used to learn $d$-dimensional node representations at each time-step. A Long Short-term Memory (LSTM) AutoEncoder (AE) is used to learn trajectory representations (similar to ~\cite{srivastava2015unsupervised}, which used this approach for videos though). The encoder-LSTM takes the sequence of $d$-dimensional node embeddings and encodes the sequence, and the decoder-LSTM reconstructs the original sequence from the embedding obtained using the encoder. Once this LSTM-AE network is completely trained, the output of the encoder-LSTM is used as the trajectory representation. This baseline algorithm was implemented using Keras ~\cite{chollet2015keras}, and trained using the Adam optimizer with the mean square error loss function. The low-dimensional embedding obtained through encoder was chosen to be of $d/2$ dimensions after empirical studies.

\subsection{\STE \ for Trajectory Classification}
\label{subsec:classi}
We studied the effectiveness of the representations of trajectories learned using the proposed methods on the task of trajectory classification (where the node's trajectory in a pre-specified time window is considered). We now describe how each of the aforementioned datasets were studied in this work for trajectory classification, along with the corresponding results. We note that in case of each of the datasets, the total number of node trajectories are divided into a 70\% training set and the remaining 30\% as a test set. An ensemble voting-based classifier (comprised of random forests and Support Vector Machines) was trained on the training set using stratified $k$-fold cross-validation with $k=10$ to finetune the hyperparameters of the classifier (this classifier was chosen after empirical studies, and is also known popularly to be a very robust classifier). The classifier with the best hyperparameters is then trained on the complete training dataset to obtain the final model, and used to study performance on test data. We use the classification accuracy as the performance metric for this task. We repeat the process five times and report the average performance in our results (Table \ref{table:accuracy}). This process is maintained for each of the considered datasets.\\

\noindent \textbf{DBLP Dataset:} To evaluate our model on the temporal graphs in the DBLP dataset, we considered non-overlapping windows of 5 years, leading to 9 such windows over the 45 graphs in this dataset. (For example, while learning the trajectory embedding of nodes at time $t_5$, we consider the graph at $t_5$ as well as the previous 4 graphs: $t_4$, $t_3$, $t_2$ and $t_1$.) A similar process is used for window of 10 years too. 

The representations of author trajectories learned using the proposed methods and baseline methods are then used for the binary classification task, where class labels are ``Expert Researcher'' and ``Interdisciplinary Researcher''. These labels are assigned to each node trajectory in the following manner. Based on the conference where each author has published his/her work, we assign one of the labels from the following to each node of a graph at each time step. Label \textit{Database}: if published in ICDE, VLDB, SIGMOD, PODS or EDBT; label \textit{Artificial Intelligence}: if published in IJCAI, AAAI, ICML, or ECML; label \textit{Data Mining}: if published in KDD, PAKDD, ICDM, PKDD or SDM; label \textit{Information Retrieval}: if published in SIGIR, WWW, ECIR or WSDM. Subsequently, the trajectory labels are determined based on the node labels assigned at each time-step. If the node label is constant throughout the considered time window, the trajectory label corresponding to that node is ``Expert Researcher''. If the node label changes within the time window, we assign the label ``Interdisciplinary Researcher'' to the author trajectory.

\STE\ has four hyper-parameters: \textit{embedding dimension:} dimension of the trajectory representation; \textit{window size:} number of time steps to consider while learning trajectory embedding; \textit{walk length:} length of random walk performed on the spatio-temporal graph; and \textit{restarts:} number of random walks starting from the same node. Table \ref{table:accuracy} shows the performance of the proposed and baseline methods in three different hyper-parameter settings. The \textit{walk length} value was set to 30, and the \textit{restarts} value to 40.\\

\noindent \textbf{Epinion Dataset:} In this dataset, when we set the hyperparameter, window size, to five, 22 windows are considered for analysis (among the 110 graphs available in this dataset), and similarly when the window size is set to 10. Category IDs are used as node labels, which are subsequently used to determine the ground truth labels for node trajectories. If a user has reviewed products from the same category during all the time steps considered within a time window, the trajectory corresponding to the user will be assigned the label ``Proficient reviewer'', indicating that the user has good knowledge about the products in the category. If the user reviews products from different categories during the considered time window, the user is assigned the label ``Versatile reviewer'', indicating that the user has expertise in assessing products from many categories. Thus, trajectory classification is modeled as a binary classification problem on this dataset, and we report the result of executing our algorithm on Epinion dataset in various hyper-parameter settings in Table \ref{table:accuracy}. Similar to the DBLP dataset, we set the hyper-parameters \textit{walk length} to 30 and \textit{restarts} to 40 for this dataset.\\

\begin{figure*}[h]
\subfloat[Histogram shows the count of authors vs number of years spent by any author in one research area before switching to other domain\label{sfig:lstmae1}]{%
  \includegraphics[height=6cm,width=.49\linewidth]{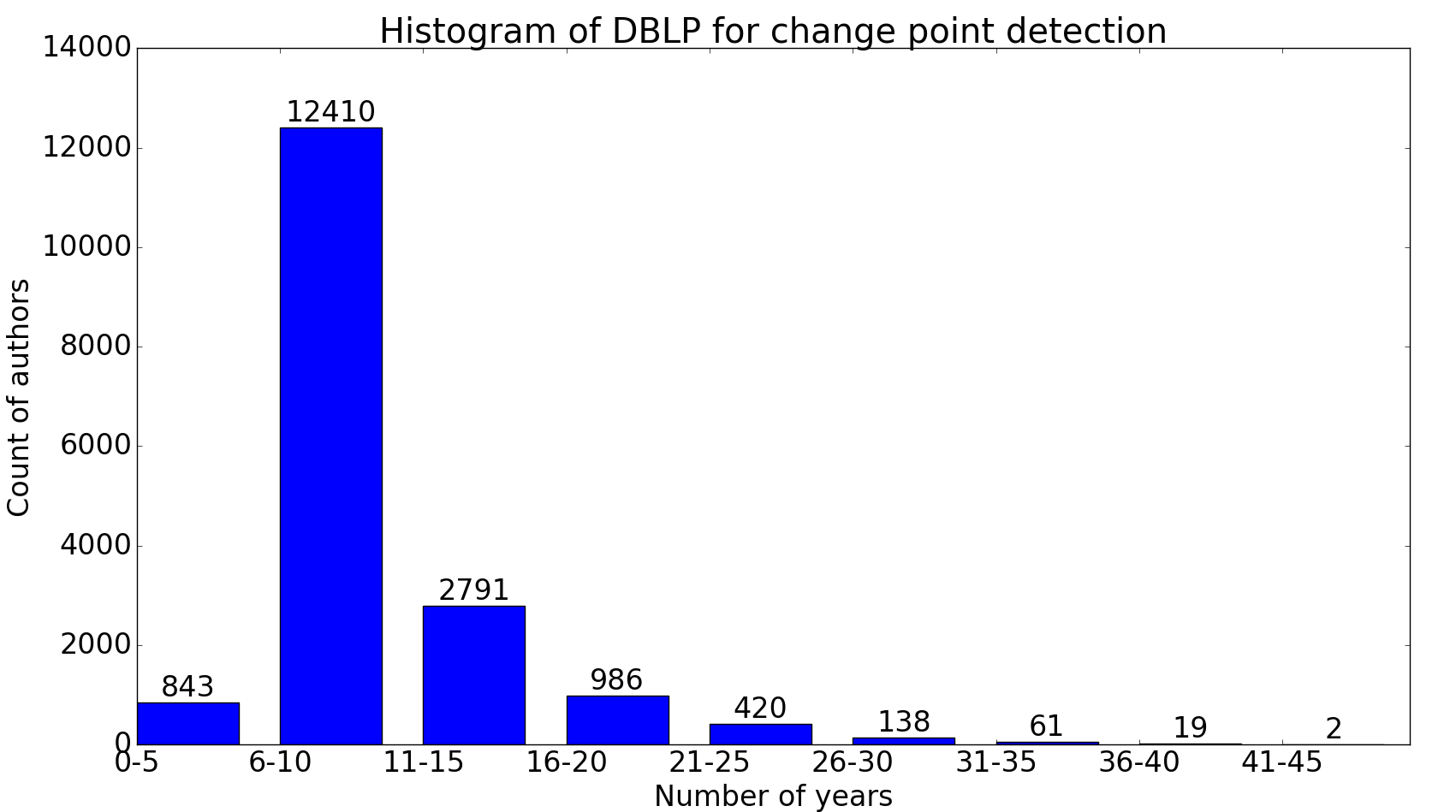}%
}\hfill
\subfloat[Histogram of number of users vs average duration spent on reviewing product of same category before switching to other category.\label{sfig:lstmae2}]{%
  \includegraphics[height=6cm,width=0.49\linewidth]{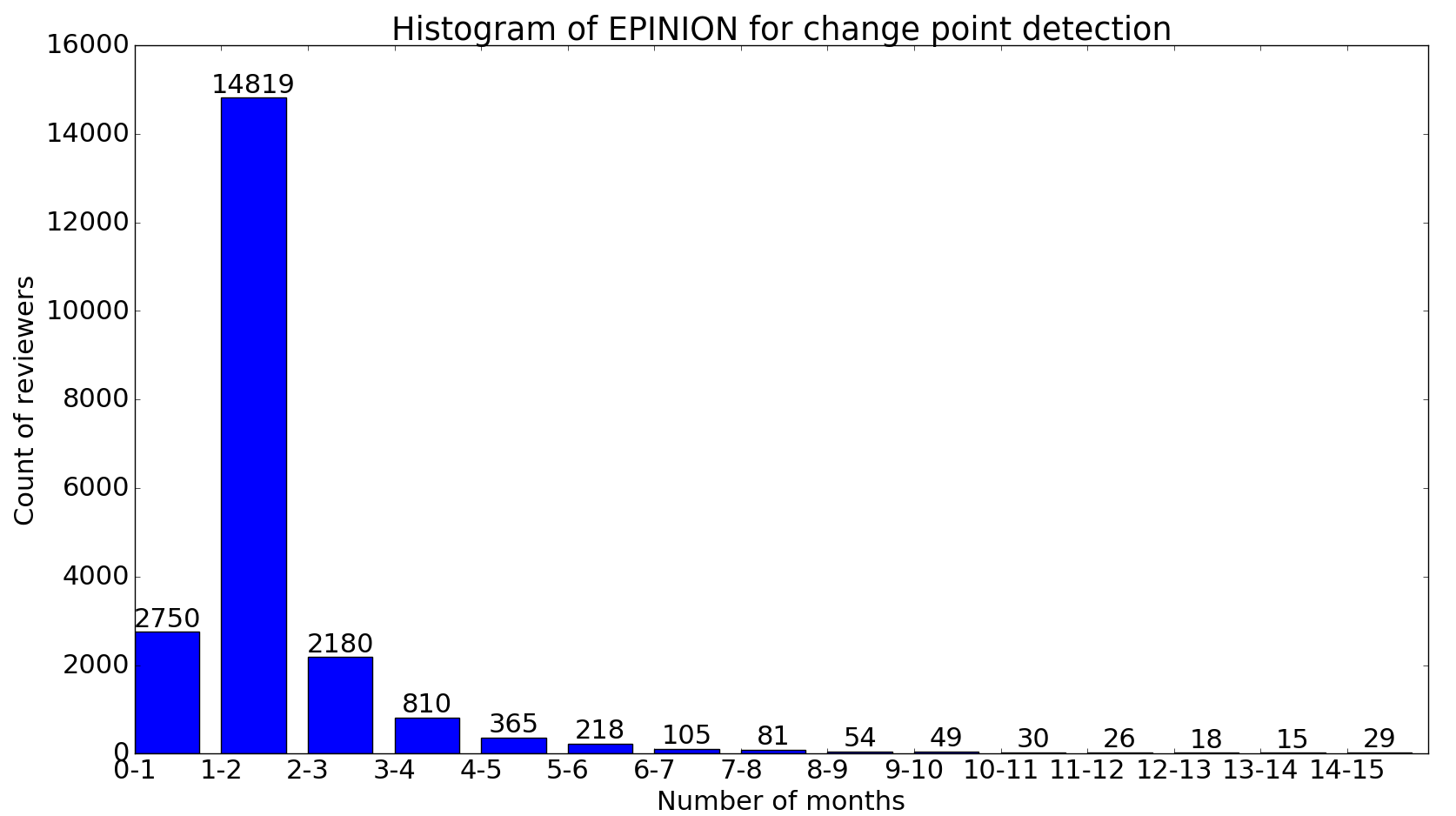}
}
\caption{Results of change point detection on the DBLP and Epinion datasets}
\label{fig:lstmaeresult}
\end{figure*}

\noindent \textbf{Ciao Dataset:} In this dataset, when we set the hyperparameter, window size, to five, 23 windows are considered for analysis (among the 115 graphs available in this dataset), and similarly when the window size is set to 10. The category ID of the product that a user reviews is considered as the node label, and ground truth trajectory labels are determined based on these node labels. Similar to the Epinion dataset, if the node labels remain constant for the entire time window, the corresponding node trajectory is assigned the label ``Experienced reviewer'', and if the user reviews products from multiple categories within the same time window, the node trajectory is assigned the label ``Multifaceted reviewer''. Binary classification is then used to validate the effectiveness of trajectory representations learned on this dataset. The \textit{walk length} and \textit{restarts} hyperparameters are chosen as on the Epinion dataset.\\

Table \ref{table:accuracy} shows the classification accuracies of the proposed and baseline methods on all the three datasets. The proposed algorithms, especially \STEtwo\, shows superior performance in most settings. In case of the CIAO dataset, the proposed algorithms show significant increase in accuracy over the baseline methods. Interestingly, the LSTM-AE method, which was introduced as a baseline method in this work, also performs better than the other two baseline methods. The results support the proposed methods in general. In particular, the improved performance of \STEtwo\ shows that having separate space and time walks provides better representations. We surmise the reason for this to be the improved coverage of temporal neighborhoods through this approach when compared to \STEone\ under the same random walk lengths.

\subsection{\STE \ for Change Point Detection}
\label{subsec:change}
In a temporal setting, analyzing change points can be critical in applications such as concept drift and anomaly detection. As the trajectory representation captures the change in behavior of nodes in temporal graph over a given time period, this makes the trajectory representations suitable features to study such change points.

In the DBLP dataset, we use such an approach to find the average time spent by an author in his/her research area before switching to another domain, thus finding the `points of change in research domain' in the author's trajectory. Specifically, we consider the trajectory labels across multiple windows of the author (as obtained from Section \ref{subsec:classi}). We study the sequence of trajectory labels and based on the pattern of occurrence of ``Interdisciplinary'' and ``Expert'' labels, we calculate the average duration of time spent in a single domain. Figure \ref{sfig:lstmae1} shows the histogram of authors who spent certain number of years in one domain before working in other research area. 
This analysis allows us to answer queries such as the minimum (or average) number of years spent before a researcher becomes interdisciplinary. For example, Figure \ref{sfig:lstmae1} shows evidently that the  average duration of time before a researcher becomes interdisciplinary is between 6-10 years, which agrees with common understanding. We also used the histogram to analyze cases of individual authors, and found the change points to be fairly accurate. This supports the effectiveness of the trajectory representations learned by the proposed methods (\STEtwo\ was used in these studies, considering its superior performance in the earlier experiments).

Similarly, analyzing change points in the Epinion dataset gives insight into the behavior of reviewers. Figure \ref{sfig:lstmae2} shows the trends for the Epinion dataset. It shows the histogram of reviewers and the number of months they review products from one single category, before switching to products from other categories. In this case, it is clear from the figure that a reviewer, on an average, writes reviews for products from one category for 1 to 2 months before switching to other categories, which matches with common understanding too. This corroborates the effectiveness of the proposed method (\STEtwo\, in this case again) for learning trajectory representations in temporal graphs.

\subsection{Arithmetic Operations on Trajectory Representations}
\label{subsec:goodness}
\begin{figure}
\includegraphics[width=9cm]{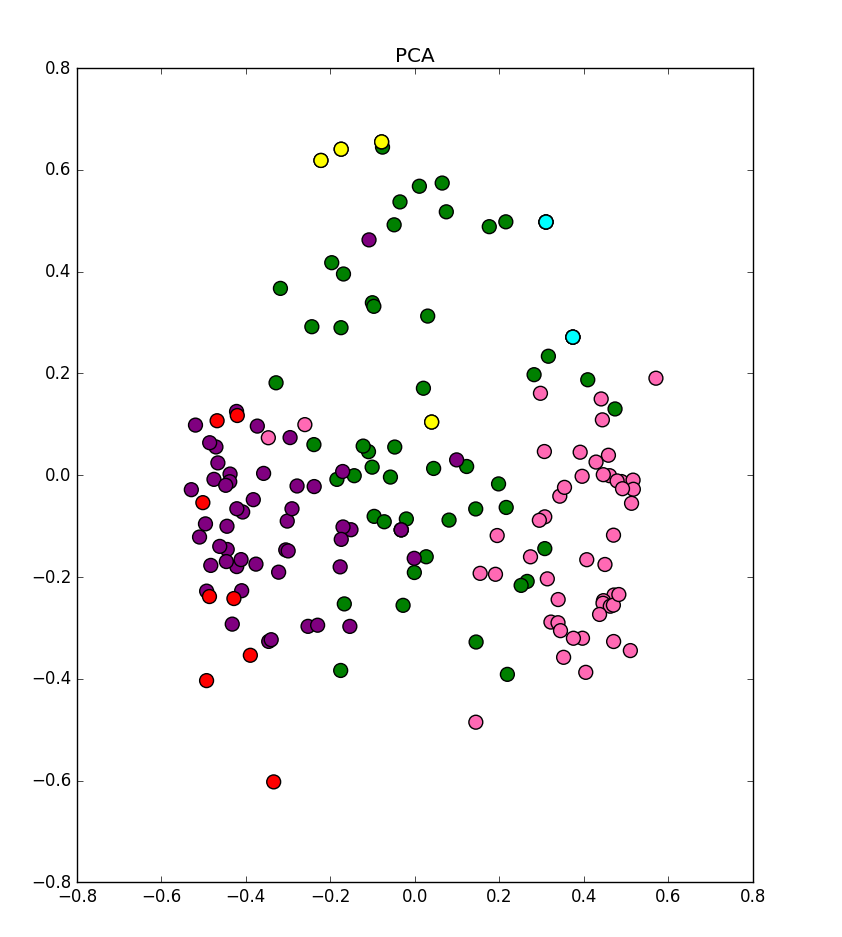}
\caption{Analyzing the goodness of trajectory embeddings of DBLP dataset(Best viewed in color). (i) Arithmetic operations are valid on trajectory embeddings. E.g. (interdisciplinary embedding)-(specialized embedding)=(other specialized embedding), i.e., taking arithmetic difference as in (AI+DB) (yellow points)-DB (cyan points)=AI (red points). (ii) Pink color cluster represents DBLP authors working in Database (DB). Violet color represents authors working in Artificial Intelligence (AI) and the green color cluster indicates the interdisciplinary authors working in both areas AI and DB.}
\label{fig:goodness}
\end{figure}

Mikolov et al.~\cite{mikolov2013distributed} demonstrated that arithmetic operations on learned word representations show a linear structure in embedding space, in the case of text processing. For example, vector(``Paris'') - vector(``France'') + vector(``England'') gives a vector representation nearest to the word embedding for London. Here, we examine whether similar behavior is shown by the learned trajectory representations. We perform this study with the representations learned on the DBLP dataset. In particular, we consider the representations for interdisciplinary authors (shown as yellow points in Figure \ref{fig:goodness}, best viewed in color) and expert authors (shown as cyan points in the figure). We then perform the following difference operation: trajectory(``interdisciplinary author'')-trajectory(``Expert author'') and examine the resulting representations (plotted in red color in the figure).

Figure \ref{fig:goodness} shows the Principal Component Analysis (PCA) plot for the result of performing arithmetic difference operation on the trajectory representations of DBLP authors, as described above. For this plot, we considered only the trajectory vectors of authors working in the \textit{Database} (DB) domain (shown in pink color on the scatter plot). We also considered the trajectory embeddings of authors who focus only on the \textit{Artificial Intelligence} (AI) domain (violet color on the scatter plot). Thus, the pink  and violet clusters represent authors with trajectory label ``Expert Researcher''. To evaluate the goodness of learned trajectory embeddings, we consider the representations for authors with labels ``Interdisciplinary Researcher''. In particular, we consider trajectories of authors who have published in both \textit{Database} and \textit{Artificial Intelligence}  (AI+DB) venues. These interdisciplinary authors are shown in green color data points on the scatter plot. Three distinct clusters, in Figure \ref{fig:goodness} indicating DB (pink), AI (violet) and AI+DB (green), validate that the proposed model \STE\ is able to learn rich trajectory embeddings for ``Expert'' authors and ``Interdisciplinary'' authors. 

To validate if the arithmetic operation: (AI+DB)-DB=AI holds for the learned trajectory embeddings, we perform the following steps. We take the trajectory vectors of a randomly picked ``Interdisciplinary'' author from the collection of vectors mentioned above. The selected ``Interdisciplinary'' points are shown in yellow color in Figure \ref{fig:goodness}. These are part of the green cluster, which represents all ``Interdisciplinary'' authors. Similarly, we pick trajectory embeddings of ``Expert'' authors from a DB background (shown in cyan color points). We then apply vector difference operation on vectors from the above two selected sets, and the resulting vectors are plotted in red color. As seen in Figure \ref{fig:goodness}, the resulting red points are near the violet points, which indicates the cluster for ``Expert'' authors in AI. This demonstrates that simple vector arithmetic on the learned trajectory representations show linear structure and hence, can be used for developing interesting applications.

\begin{figure}
\includegraphics[width=9cm,height=9cm]{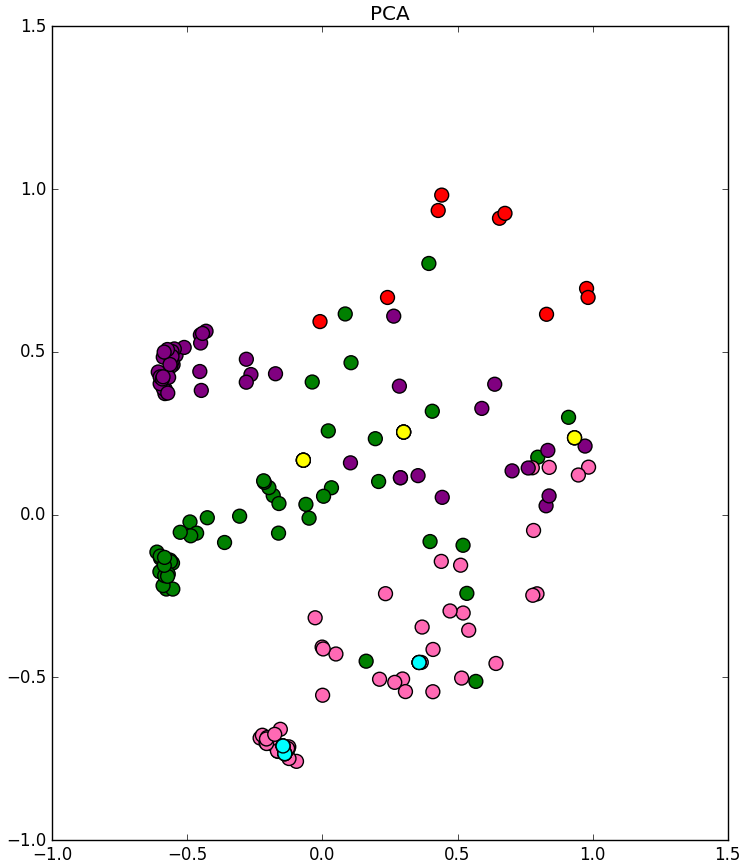}
\caption{Analyzing the goodness of trajectory embeddings of Epinion dataset (Best viewed in color). (i) Arithmetic difference of users' trajectory who review products from category (Movies+Books) (yellow points)-Books (cyan points)= Movies category (red points). (ii) Pink color cluster represents users who review Books. Violet color represents users who review Movies and the green color cluster indicates the versatile users who review products from both categories Movies and Books.}
\label{fig:goodness1}
\end{figure}

Figure \ref{fig:goodness1} shows the results of similar vector operation on Epinion dataset. We considered trajectory representation of users who review products from category \textit{Movies} and \textit{Books} as ``Proficient reviewer'', represented by violet and pink clusters respectively in figure \ref{fig:goodness1}. The trajectory representation of users reviewing products from both categories \textit{Movies} and \textit{Books} as ``Versatile reviewer'', represented by green cluster. We randomly select trajectory embedding of ``Versatile reviewer'', the yellow points and ``Proficient reviewer'', the cyan points, and then perform vector subtraction. The resulting points are plotted in red color. We can conclude from \ref{fig:goodness1} that these points are near violet cluster which represents \textit{Movies} cluster, validating that the learned trajectory representations show linear structure


\section{Conclusion}\label{sec:conc}

In this paper, we proposed \STE\, a novel approach to learn trajectory representations in temporal graphs, by capturing change in nodes' behavior over time. We presented two algorithms to learn such trajectory representations: \STEone\ learns the representations by considering the present neighbors and past neighbors together, and \STEtwo\ models it as two subproblems, solving each separately before combining the results to get the final trajectory embedding. We validated the effectiveness of the learned trajectory representations on three real-world datasets: DBLP, Epinion and Ciao. We compare the results against  three baseline algorithms, and show that the proposed \STE\ outperforms baseline methods in various settings.

Furthermore, we demonstrated that simple arithmetic operations are valid on the trajectory representations learned through the proposed methods, i.e., the vector difference: trajectory(``interdisciplinary author'')- trajectory(``specialized author'')=trajectory(``other specialized author''). We also examined the usefulness of trajectory embeddings in change point detection. Each of these experimental studies showed the effectiveness of trajectory representations learned by \STE.


\bibliographystyle{ACM-Reference-Format}
\bibliography{references} 


\begin{thebibliography}{00}


\ifx \showCODEN    \undefined \def \showCODEN     #1{\unskip}     \fi
\ifx \showDOI      \undefined \def \showDOI       #1{#1}\fi
\ifx \showISBNx    \undefined \def \showISBNx     #1{\unskip}     \fi
\ifx \showISBNxiii \undefined \def \showISBNxiii  #1{\unskip}     \fi
\ifx \showISSN     \undefined \def \showISSN      #1{\unskip}     \fi
\ifx \showLCCN     \undefined \def \showLCCN      #1{\unskip}     \fi
\ifx \shownote     \undefined \def \shownote      #1{#1}          \fi
\ifx \showarticletitle \undefined \def \showarticletitle #1{#1}   \fi
\ifx \showURL      \undefined \def \showURL       {\relax}        \fi
\providecommand\bibfield[2]{#2}
\providecommand\bibinfo[2]{#2}
\providecommand\natexlab[1]{#1}
\providecommand\showeprint[2][]{arXiv:#2}

\bibitem[\protect\citeauthoryear{Aggarwal and Subbian}{Aggarwal and
  Subbian}{2014}]%
        {aggarwal2014evolutionary}
\bibfield{author}{\bibinfo{person}{Charu Aggarwal} {and}
  \bibinfo{person}{Karthik Subbian}.} \bibinfo{year}{2014}\natexlab{}.
\newblock \showarticletitle{Evolutionary network analysis: A survey}.
\newblock \bibinfo{journal}{{\em ACM Computing Surveys (CSUR)\/}}
  \bibinfo{volume}{47}, \bibinfo{number}{1} (\bibinfo{year}{2014}),
  \bibinfo{pages}{10}.
\newblock


\bibitem[\protect\citeauthoryear{Aggarwal}{Aggarwal}{2011}]%
        {aggarwal2011introduction}
\bibfield{author}{\bibinfo{person}{Charu~C Aggarwal}.}
  \bibinfo{year}{2011}\natexlab{}.
\newblock \showarticletitle{An introduction to social network data analytics}.
\newblock \bibinfo{journal}{{\em Social network data analytics\/}}
  (\bibinfo{year}{2011}), \bibinfo{pages}{1--15}.
\newblock


\bibitem[\protect\citeauthoryear{Aggarwal and Li}{Aggarwal and Li}{2011}]%
        {aggarwal2011node}
\bibfield{author}{\bibinfo{person}{Charu~C Aggarwal} {and} \bibinfo{person}{Nan
  Li}.} \bibinfo{year}{2011}\natexlab{}.
\newblock \showarticletitle{On node classification in dynamic content-based
  networks}. In \bibinfo{booktitle}{{\em Proceedings of the 2011 SIAM
  International Conference on Data Mining}}. SIAM, \bibinfo{pages}{355--366}.
\newblock


\bibitem[\protect\citeauthoryear{Akoglu, Tong, and Koutra}{Akoglu
  et~al\mbox{.}}{2015}]%
        {akoglu2015graph}
\bibfield{author}{\bibinfo{person}{Leman Akoglu}, \bibinfo{person}{Hanghang
  Tong}, {and} \bibinfo{person}{Danai Koutra}.}
  \bibinfo{year}{2015}\natexlab{}.
\newblock \showarticletitle{Graph based anomaly detection and description: a
  survey}.
\newblock \bibinfo{journal}{{\em Data Mining and Knowledge Discovery\/}}
  \bibinfo{volume}{29}, \bibinfo{number}{3} (\bibinfo{year}{2015}),
  \bibinfo{pages}{626--688}.
\newblock


\bibitem[\protect\citeauthoryear{Cao, Lu, and Xu}{Cao et~al\mbox{.}}{2015}]%
        {cao2015grarep}
\bibfield{author}{\bibinfo{person}{Shaosheng Cao}, \bibinfo{person}{Wei Lu},
  {and} \bibinfo{person}{Qiongkai Xu}.} \bibinfo{year}{2015}\natexlab{}.
\newblock \showarticletitle{Grarep: Learning graph representations with global
  structural information}. In \bibinfo{booktitle}{{\em Proceedings of the 24th
  ACM International on Conference on Information and Knowledge Management}}.
  ACM, \bibinfo{pages}{891--900}.
\newblock


\bibitem[\protect\citeauthoryear{Cao, Lu, and Xu}{Cao et~al\mbox{.}}{2016}]%
        {cao2016deep}
\bibfield{author}{\bibinfo{person}{Shaosheng Cao}, \bibinfo{person}{Wei Lu},
  {and} \bibinfo{person}{Qiongkai Xu}.} \bibinfo{year}{2016}\natexlab{}.
\newblock \showarticletitle{Deep Neural Networks for Learning Graph
  Representations.}. In \bibinfo{booktitle}{{\em AAAI}}.
  \bibinfo{pages}{1145--1152}.
\newblock


\bibitem[\protect\citeauthoryear{Chang, Han, Tang, Qi, Aggarwal, and
  Huang}{Chang et~al\mbox{.}}{2015}]%
        {chang2015heterogeneous}
\bibfield{author}{\bibinfo{person}{Shiyu Chang}, \bibinfo{person}{Wei Han},
  \bibinfo{person}{Jiliang Tang}, \bibinfo{person}{Guo-Jun Qi},
  \bibinfo{person}{Charu~C Aggarwal}, {and} \bibinfo{person}{Thomas~S Huang}.}
  \bibinfo{year}{2015}\natexlab{}.
\newblock \showarticletitle{Heterogeneous network embedding via deep
  architectures}. In \bibinfo{booktitle}{{\em Proceedings of the 21th ACM
  SIGKDD International Conference on Knowledge Discovery and Data Mining}}.
  ACM, \bibinfo{pages}{119--128}.
\newblock


\bibitem[\protect\citeauthoryear{Chollet et~al\mbox{.}}{Chollet
  et~al\mbox{.}}{2015}]%
        {chollet2015keras}
\bibfield{author}{\bibinfo{person}{Fran\c{c}ois Chollet} {et~al\mbox{.}}}
  \bibinfo{year}{2015}\natexlab{}.
\newblock \bibinfo{title}{Keras}.
\newblock \bibinfo{howpublished}{\url{https://github.com/fchollet/keras}}.
  (\bibinfo{year}{2015}).
\newblock


\bibitem[\protect\citeauthoryear{Getoor and Diehl}{Getoor and Diehl}{2005}]%
        {getoor2005link}
\bibfield{author}{\bibinfo{person}{Lise Getoor} {and}
  \bibinfo{person}{Christopher~P Diehl}.} \bibinfo{year}{2005}\natexlab{}.
\newblock \showarticletitle{Link mining: a survey}.
\newblock \bibinfo{journal}{{\em Acm Sigkdd Explorations Newsletter\/}}
  \bibinfo{volume}{7}, \bibinfo{number}{2} (\bibinfo{year}{2005}),
  \bibinfo{pages}{3--12}.
\newblock


\bibitem[\protect\citeauthoryear{Goyal and Ferrara}{Goyal and Ferrara}{2017}]%
        {goyal2017graph}
\bibfield{author}{\bibinfo{person}{Palash Goyal} {and} \bibinfo{person}{Emilio
  Ferrara}.} \bibinfo{year}{2017}\natexlab{}.
\newblock \showarticletitle{Graph Embedding Techniques, Applications, and
  Performance: A Survey}.
\newblock \bibinfo{journal}{{\em arXiv preprint arXiv:1705.02801\/}}
  (\bibinfo{year}{2017}).
\newblock


\bibitem[\protect\citeauthoryear{Grover and Leskovec}{Grover and
  Leskovec}{2016}]%
        {grover2016node2vec}
\bibfield{author}{\bibinfo{person}{Aditya Grover} {and} \bibinfo{person}{Jure
  Leskovec}.} \bibinfo{year}{2016}\natexlab{}.
\newblock \showarticletitle{node2vec: Scalable feature learning for networks}.
  In \bibinfo{booktitle}{{\em Proceedings of the 22nd ACM SIGKDD International
  Conference on Knowledge Discovery and Data Mining}}. ACM,
  \bibinfo{pages}{855--864}.
\newblock


\bibitem[\protect\citeauthoryear{Gupta, Gao, Aggarwal, and Han}{Gupta
  et~al\mbox{.}}{2014}]%
        {gupta2014outlier}
\bibfield{author}{\bibinfo{person}{Manish Gupta}, \bibinfo{person}{Jing Gao},
  \bibinfo{person}{Charu Aggarwal}, {and} \bibinfo{person}{Jiawei Han}.}
  \bibinfo{year}{2014}\natexlab{}.
\newblock \showarticletitle{Outlier detection for temporal data}.
\newblock \bibinfo{journal}{{\em Synthesis Lectures on Data Mining and
  Knowledge Discovery\/}} \bibinfo{volume}{5}, \bibinfo{number}{1}
  (\bibinfo{year}{2014}), \bibinfo{pages}{1--129}.
\newblock


\bibitem[\protect\citeauthoryear{Hamilton, Ying, and Leskovec}{Hamilton
  et~al\mbox{.}}{2017a}]%
        {hamilton2017inductive}
\bibfield{author}{\bibinfo{person}{William~L Hamilton}, \bibinfo{person}{Rex
  Ying}, {and} \bibinfo{person}{Jure Leskovec}.}
  \bibinfo{year}{2017}\natexlab{a}.
\newblock \showarticletitle{Inductive Representation Learning on Large Graphs}.
\newblock \bibinfo{journal}{{\em arXiv preprint arXiv:1706.02216\/}}
  (\bibinfo{year}{2017}).
\newblock


\bibitem[\protect\citeauthoryear{Hamilton, Ying, and Leskovec}{Hamilton
  et~al\mbox{.}}{2017b}]%
        {hamilton2017representation}
\bibfield{author}{\bibinfo{person}{William~L Hamilton}, \bibinfo{person}{Rex
  Ying}, {and} \bibinfo{person}{Jure Leskovec}.}
  \bibinfo{year}{2017}\natexlab{b}.
\newblock \showarticletitle{Representation Learning on Graphs: Methods and
  Applications}.
\newblock \bibinfo{journal}{{\em arXiv preprint arXiv:1709.05584\/}}
  (\bibinfo{year}{2017}).
\newblock


\bibitem[\protect\citeauthoryear{Li, Dani, Hu, Tang, Chang, and Liu}{Li
  et~al\mbox{.}}{2017}]%
        {li2017attributed}
\bibfield{author}{\bibinfo{person}{Jundong Li}, \bibinfo{person}{Harsh Dani},
  \bibinfo{person}{Xia Hu}, \bibinfo{person}{Jiliang Tang}, \bibinfo{person}{Yi
  Chang}, {and} \bibinfo{person}{Huan Liu}.} \bibinfo{year}{2017}\natexlab{}.
\newblock \showarticletitle{Attributed Network Embedding for Learning in a
  Dynamic Environment}.
\newblock \bibinfo{journal}{{\em arXiv preprint arXiv:1706.01860\/}}
  (\bibinfo{year}{2017}).
\newblock


\bibitem[\protect\citeauthoryear{Malliaros and Vazirgiannis}{Malliaros and
  Vazirgiannis}{2013}]%
        {malliaros2013clustering}
\bibfield{author}{\bibinfo{person}{Fragkiskos~D Malliaros} {and}
  \bibinfo{person}{Michalis Vazirgiannis}.} \bibinfo{year}{2013}\natexlab{}.
\newblock \showarticletitle{Clustering and community detection in directed
  networks: A survey}.
\newblock \bibinfo{journal}{{\em Physics Reports\/}} \bibinfo{volume}{533},
  \bibinfo{number}{4} (\bibinfo{year}{2013}), \bibinfo{pages}{95--142}.
\newblock


\bibitem[\protect\citeauthoryear{Mikolov, Chen, Corrado, and Dean}{Mikolov
  et~al\mbox{.}}{2013a}]%
        {mikolov2013efficient}
\bibfield{author}{\bibinfo{person}{Tomas Mikolov}, \bibinfo{person}{Kai Chen},
  \bibinfo{person}{Greg Corrado}, {and} \bibinfo{person}{Jeffrey Dean}.}
  \bibinfo{year}{2013}\natexlab{a}.
\newblock \showarticletitle{Efficient estimation of word representations in
  vector space}.
\newblock \bibinfo{journal}{{\em arXiv preprint arXiv:1301.3781\/}}
  (\bibinfo{year}{2013}).
\newblock


\bibitem[\protect\citeauthoryear{Mikolov, Sutskever, Chen, Corrado, and
  Dean}{Mikolov et~al\mbox{.}}{2013b}]%
        {mikolov2013distributed}
\bibfield{author}{\bibinfo{person}{Tomas Mikolov}, \bibinfo{person}{Ilya
  Sutskever}, \bibinfo{person}{Kai Chen}, \bibinfo{person}{Greg~S Corrado},
  {and} \bibinfo{person}{Jeff Dean}.} \bibinfo{year}{2013}\natexlab{b}.
\newblock \showarticletitle{Distributed representations of words and phrases
  and their compositionality}. In \bibinfo{booktitle}{{\em Advances in neural
  information processing systems}}. \bibinfo{pages}{3111--3119}.
\newblock


\bibitem[\protect\citeauthoryear{Perozzi, Al-Rfou, and Skiena}{Perozzi
  et~al\mbox{.}}{2014}]%
        {Perozzi:2014:DOL:2623330.2623732}
\bibfield{author}{\bibinfo{person}{Bryan Perozzi}, \bibinfo{person}{Rami
  Al-Rfou}, {and} \bibinfo{person}{Steven Skiena}.}
  \bibinfo{year}{2014}\natexlab{}.
\newblock \showarticletitle{DeepWalk: Online Learning of Social
  Representations}. In \bibinfo{booktitle}{{\em Proceedings of the 20th ACM
  SIGKDD International Conference on Knowledge Discovery and Data Mining}} {\em
  (\bibinfo{series}{KDD '14})}. \bibinfo{publisher}{ACM}, \bibinfo{address}{New
  York, NY, USA}, \bibinfo{pages}{701--710}.
\newblock
\showISBNx{978-1-4503-2956-9}
\showDOI{%
\url{https://doi.org/10.1145/2623330.2623732}}


\bibitem[\protect\citeauthoryear{Sarkar, Chakrabarti, and Jordan}{Sarkar
  et~al\mbox{.}}{2012}]%
        {sarkar2012nonparametric}
\bibfield{author}{\bibinfo{person}{Purnamrita Sarkar},
  \bibinfo{person}{Deepayan Chakrabarti}, {and} \bibinfo{person}{Michael
  Jordan}.} \bibinfo{year}{2012}\natexlab{}.
\newblock \showarticletitle{Nonparametric link prediction in dynamic networks}.
\newblock \bibinfo{journal}{{\em arXiv preprint arXiv:1206.6394\/}}
  (\bibinfo{year}{2012}).
\newblock


\bibitem[\protect\citeauthoryear{Srivastava, Mansimov, and
  Salakhutdinov}{Srivastava et~al\mbox{.}}{2015}]%
        {srivastava2015unsupervised}
\bibfield{author}{\bibinfo{person}{Nitish Srivastava}, \bibinfo{person}{Elman
  Mansimov}, {and} \bibinfo{person}{Ruslan Salakhutdinov}.}
  \bibinfo{year}{2015}\natexlab{}.
\newblock \showarticletitle{Unsupervised Learning of Video Representations
  using LSTMs.}. In \bibinfo{booktitle}{{\em ICML}}. \bibinfo{pages}{843--852}.
\newblock


\bibitem[\protect\citeauthoryear{Tang, Gao, and Liu}{Tang
  et~al\mbox{.}}{2012a}]%
        {tang-etal12a}
\bibfield{author}{\bibinfo{person}{J. Tang}, \bibinfo{person}{H. Gao}, {and}
  \bibinfo{person}{H. Liu}.} \bibinfo{year}{2012}\natexlab{a}.
\newblock \showarticletitle{m{T}rust: {D}iscerning multi-faceted trust in a
  connected world}. In \bibinfo{booktitle}{{\em Proceedings of the fifth ACM
  international conference on Web search and data mining}}. ACM,
  \bibinfo{pages}{93--102}.
\newblock


\bibitem[\protect\citeauthoryear{Tang, Gao, Liu, and Das~Sarma}{Tang
  et~al\mbox{.}}{2012b}]%
        {tang-etal12b}
\bibfield{author}{\bibinfo{person}{J. Tang}, \bibinfo{person}{H. Gao},
  \bibinfo{person}{H. Liu}, {and} \bibinfo{person}{A. Das~Sarma}.}
  \bibinfo{year}{2012}\natexlab{b}.
\newblock \bibinfo{title}{e{T}rust: {U}nderstanding trust evolution in an
  online world}.
\newblock   (\bibinfo{year}{2012}), \bibinfo{numpages}{253--261}~pages.
\newblock


\bibitem[\protect\citeauthoryear{Tang, Qu, Wang, Zhang, Yan, and Mei}{Tang
  et~al\mbox{.}}{2015}]%
        {tang2015line}
\bibfield{author}{\bibinfo{person}{Jian Tang}, \bibinfo{person}{Meng Qu},
  \bibinfo{person}{Mingzhe Wang}, \bibinfo{person}{Ming Zhang},
  \bibinfo{person}{Jun Yan}, {and} \bibinfo{person}{Qiaozhu Mei}.}
  \bibinfo{year}{2015}\natexlab{}.
\newblock \showarticletitle{Line: Large-scale information network embedding}.
  In \bibinfo{booktitle}{{\em Proceedings of the 24th International Conference
  on World Wide Web}}. ACM, \bibinfo{pages}{1067--1077}.
\newblock


\bibitem[\protect\citeauthoryear{Tang, Liu, Zhang, and Nazeri}{Tang
  et~al\mbox{.}}{2008}]%
        {tang2008community}
\bibfield{author}{\bibinfo{person}{Lei Tang}, \bibinfo{person}{Huan Liu},
  \bibinfo{person}{Jianping Zhang}, {and} \bibinfo{person}{Zohreh Nazeri}.}
  \bibinfo{year}{2008}\natexlab{}.
\newblock \showarticletitle{Community evolution in dynamic multi-mode
  networks}. In \bibinfo{booktitle}{{\em Proceedings of the 14th ACM SIGKDD
  international conference on Knowledge discovery and data mining}}. ACM,
  \bibinfo{pages}{677--685}.
\newblock


\end{thebibliography}

\end{document}